\documentstyle[preprint,eqsecnum,aps,epsfig]{revtex}
\tightenlines

\newcommand{\sub}[2]{\mbox{${\mbox{#1}}_{\mbox{#2}}$}}
\newcommand{\super}[2]{\mbox{${\mbox{#1}}^{\mbox{\footnotesize #2}}$}}

\begin{document}
\title{Quantum Annealing of a Disordered Spin System}
\author{J. Brooke, D. Bitko, T. F. Rosenbaum}
\address{
The James Franck Institute and Department of Physics, The University
of Chicago\\ Chicago, Illinois  60637
}
\author{G. Aeppli}
\address{
AT\&T Bell Laboratories, 4 Independence Way, Princeton, New Jersey  07974
}
\maketitle

\begin{abstract}
Traditional simulated annealing utilizes thermal fluctuations for
convergence in optimization problems.  Quantum tunneling provides a
different mechanism for moving between states, with the potential for
reduced time scales.  We compare thermal and quantum annealing in a
model disordered Ising magnet,
Li\sub{Ho}{0.44}\sub{Y}{0.56}\sub{F}{4}, where the effects of quantum
mechanics can be tuned in the laboratory by varying a magnetic field
applied transverse to the Ising axis.  The results indicate that
quantum annealing indeed hastens convergence to the optimum state.
\end{abstract}

\begin{figure}[p!]
  \centering \epsfig{file=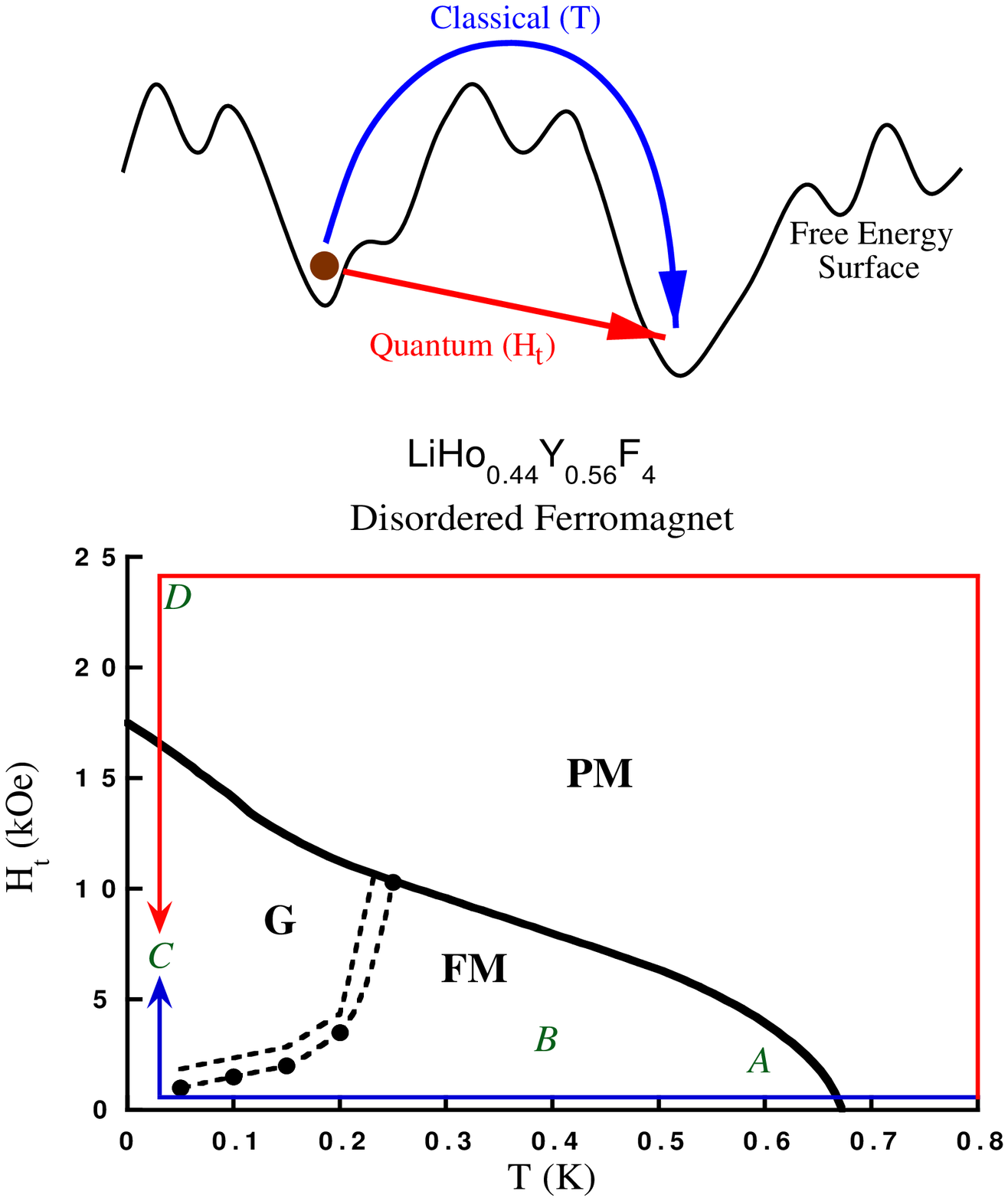, width=5in}
  \caption[Phase diagram showing quantum and classical annealing protocols.]{ 
Phase diagram showing quantum and classical annealing protocols.  The
quantum (red) and classical (blue) annealing protocols can provide
different pathways through the free energy surface.  PM = paramagnet, 
FM = ferromagnet, and G = glass, with the dashed line demarcating a dynamic 
crossover between manifestly FM and G regimes.    A-D refer to
specific points in the $H_t-T$ plane discussed in the text.}
  \label{QC1}
\end{figure}

In their presentation of simulated annealing, Kirkpatrick, Gelatt and
Vecchi~\cite{a:SimAnneal} described a fundamental connection between
statistical mechanics and combinatorial optimization.  Complex systems
subject to conflicting constraints, from the traveling salesman
problem and circuit design on one hand to spin glasses and protein
folding on the other, are difficult to solve because of the vast
number of nearly degenerate solutions.  The introduction of a variable
``temperature'' permits the simulation to naturally subdivide a
problem by energy scale, and as the temperature approaches zero the
system settles into a local minimum (Fig.~\ref{QC1}) that should be
comparable to the ground state of the system.

If the settling is performed sufficiently slowly, then the minimum is
guaranteed to be the ground state~\cite{Geman_GroundState}.  However,
complex systems with many degrees of freedom may require impractically
long annealing schedules to find the true lowest energy configuration.
If barriers between adjacent energy minima are very high yet
sufficiently narrow, it may be that quantum tunneling is a more
effective means at energy minimization than pure thermal processes,
with the potential to hasten convergence to the ground state.

As an example, consider the two-state spin system (``up'' and ``down'')
used to introduce tunneling in quantum mechanics.  The application of a
magnetic field perpendicular to the up-down axis induces off-diagonal terms
in the Hamiltonian, and enables tunneling between the two measured states.
The assembly of a macroscopic number of such quantum spins on a
lattice represents Feynman's original concept~\cite{Feynman_SimPhys}
of a quantum mechanical computer.  Information at the inputs (the
original spin state of the system) undergoes a series of quantum
mechanical operations, with the final set of ones and zeroes read at
the outputs (the optimized, low energy spin state).  Our experiment investigated
a nontrivial optimization problem in 
statistical mechanics, namely that of finding the ground state for a
ferromagnet with a certain proportion of randomly inserted
antiferromagnetic bonds (which favor antiparallel alignment of spins),
and whether this problem can be solved more rapidly by quantum
annealing than by classical thermal annealing.  We started from the
disordered, paramagnetic, high-temperature state in the dipolar-coupled Ising ferromagnet
Li\sub{Ho}{x}\sub{Y}{{\footnotesize 1}-x}\sub{F}{\footnotesize 4}, 
and read out the optimized low-temperature state using
conventional magnetic susceptometry.

The Ising magnet Li\sub{Ho}{x}\sub{Y}{{\footnotesize 1}-x}\sub{F}{\footnotesize 4}
in a transverse magnetic field $H_t$ is the experimental realization of the simplest 
quantum spin model.  The corresponding Hamiltonian (${\mathcal H}$) is 
\begin{equation}
  {\mathcal H} = -\sum_{i,j}^{N}J_{i,j} \sigma_i^z \sigma_j^z
                 -\Gamma \sum_i^N \sigma_i^x, \label{eq:Hamiltonian}
\end{equation}
where the $\sigma$'s are Pauli spin matrices, the $J_{i,j}$'s are
longitudinal couplings, and $\Gamma$ is a transverse field.  Given that the
commutator $[{\mathcal H},\sigma^z]$ is finite when $\Gamma \ne 0$, $\sigma^z$ is no longer conserved
and zero-point (quantum)
fluctuations appear.  These fluctuations
increase with $\Gamma$, which tunes an order-disorder transition at
$T=0$.  Our experiments are very different from traditional experiments on disordered 
magnets where fields are applied parallel to an easy direction for 
magnetization~\cite{Mydosh_SG,Reich}. 
 In particular, for Ising systems such as ferromagnetic 
Li\sub{Ho}{x}\sub{Y}{{\footnotesize 1}-x}\sub{F}{\footnotesize 4}, 
a longitudinal field simply polarizes 
the spins at all $T$ and removes the possibility of a ferromagnetic phase transition.  By contrast, 
the transverse field $\Gamma$ is not conjugate to the order parameter and retains a true phase transition.  
With this system we can directly compare the efficiency of quantum tunneling to classical thermal 
relaxation in finding the minima of a complicated energy landscape consisting of $N \sim 10^{23}$ spins.

In our experiments, the magnetic field $H_{t}$ was applied
perpendicular to the Ising axis (crystalline $c$ axis) for the Ho spins.  At low
temperatures ($T < 1$ K), the only \super{Ho}{3+} crystal field state
which is appreciably populated is the ($H_{t}=0$) ground-state
doublet, which can be split in continuous fashion with great precision
by the laboratory field $H_{t}$~\cite{Wu_Glass,Wu_SG,Bitko_QC,Hansen_CrysField}.  The
splitting $\Gamma$ plays the role of the transverse field in Eq.~\ref{eq:Hamiltonian},
whereas the doublet plays the role of the spin-1/2 eigenstates.  For the
present experiment, we selected a single crystal with 56\% substitution of
magnetically inert Y for Ho.  The random dilution of magnetic by nonmagnetic
ions yields couplings between the magnetic ions of effectively random 
sign~\cite{Aharony,Reich_Rev},
which make the search for the ground state difficult in the rare earth lithium
fluorides.  The crystal was ground to a needle-like cylinder of aspect ratio 
three to minimize demagnetization effects.  We suspended the cylinder
from the mixing chamber of a helium dilution refrigerator
inside the bore of an 80 kOe superconducting magnet, with the field
direction oriented along the crystal $a$-axis (within \super{5}{o}),
perpendicular to the Ising axis (within \super{0.5}{o}).  A trim coil
oriented along the Ising direction nulled any unwanted longitudinal
field component.  The protocol for comparing equilibration due to
quantum tunneling and thermal hopping was straightforward.  We
annealed using both methods to the same point in the
temperature-transverse field plane, and then measured the complete ac
susceptibility, $\chi(f) = \chi'(f)+i\chi''(f)$, along the Ising axis
with a standard gradiometer configuration and spectrum analyzer.
Typical measuring frequencies ($f$) ranged between $10^0$ and $10^5$ Hz.

\begin{figure}[p!]
  \centering \epsfig{file=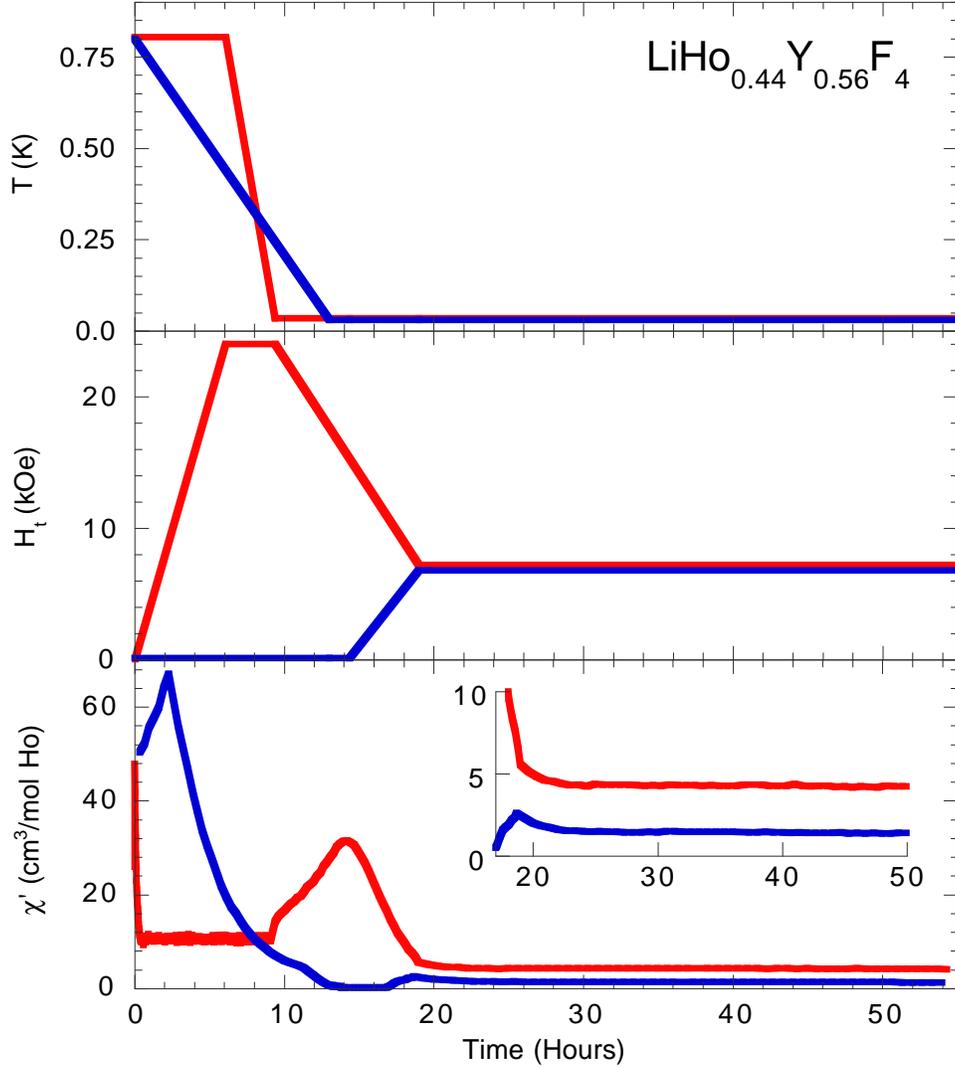, width=5in}
  \caption
[Time evolution of the real component of susceptibility following annealing
histories.]
{Time evolution of the real part of the magnetic susceptibility, $\chi'$,
following classical (blue) and quantum (red) annealing histories at $f=15$ Hz.
Although the end
point at $T=0.03$~K and $H_t=7.2$~kOe is identical, the long-time state
of the system is different.  The demagnetization limit for the sample geometry is
$\chi' = 67$ \super{cm}{3}/mol Ho.}
  \label{QC2}
\end{figure}

At low temperatures and large transverse fields, the mix of quantum mechanics 
and disorder converts 
Li\sub{Ho}{\footnotesize 0.44}\sub{Y}{\footnotesize 0.56}\sub{F}{\footnotesize 4} 
from a classical ferromagnet with Curie temperature 
${\mathrm T}_{\mathrm c} ({\mathrm H}_{\mathrm t}=0) = 0.673$~K to a magnet 
with glassy, history-dependent behavior~\cite{Bitko_Thesis}.  
The $H_t-T$ phase diagram as well as two cooling protocols are
shown in Fig.~\ref{QC1}.  The classical route (blue) crosses the phase
boundary in zero transverse field, decreasing $T$ from 0.800~K
to 0.030~K, and only then raising $H_t$ to 7.2~kOe, whereas the quantum
route (red), cools to 0.030~K in large transverse field (24~kOe),
proceeding through the order-disorder transition with finite
$H_t$ and significant tunneling potential to the same nominal end
point.  The variation of $T$, $H_t$, and the real part
of the magnetic susceptibility, $\chi'(f=15\ \mathrm{Hz})$, with time
$t$ are compared for the thermal and the quantum computation
(Fig.~\ref{QC2}).  Remarkably, the magnetic susceptibility of the
system at the identical place in the $H_t - T$ plane
arrives at a different value depending on the annealing protocol.
This difference survives to long times, at least on the order days.

\begin{figure}[p!]
  \centering \epsfig{file=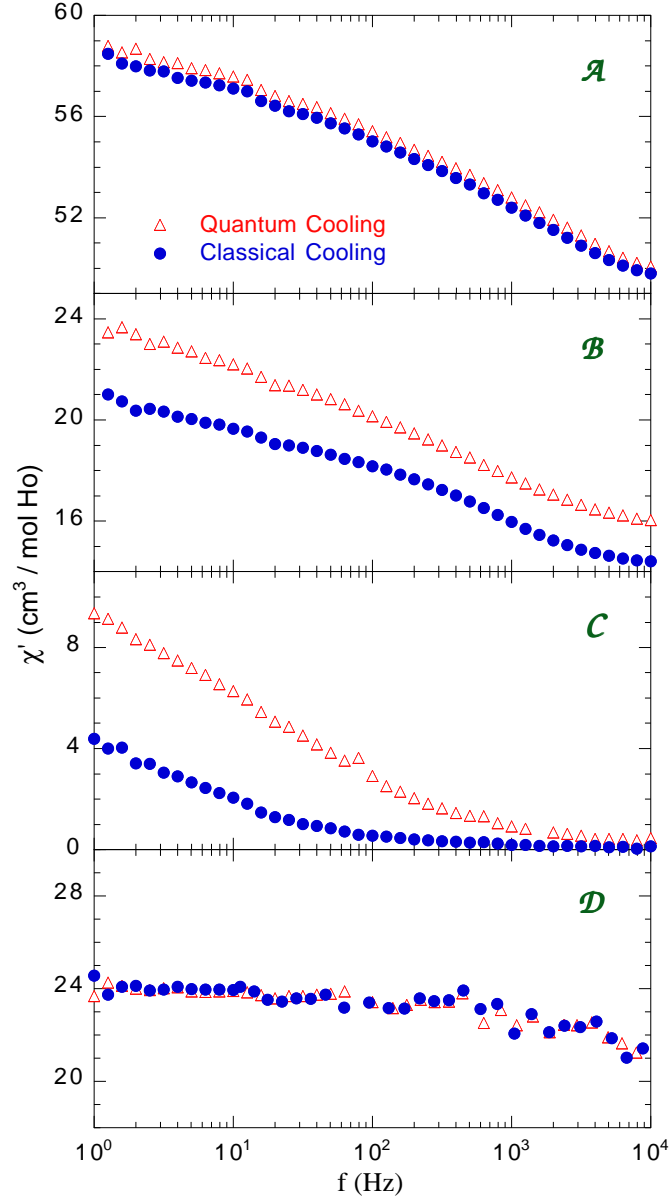, width=3.5in}
  \caption
[Spectroscopy of Li\sub{Ho}{0.44}\sub{Y}{0.56}\sub{F}{4} after quantum
and classical cooling.]
{Spectroscopy of Li\sub{Ho}{0.44}\sub{Y}{0.56}\sub{F}{4} at the points
A-D in Fig.~\ref{QC1} after both quantum and classical computations.
While the spectra begin together (A) in the classical ferromagnet,
they start to diverge as T is lowered (B), until deep in the glassy
phase (C) they exhibit widely different time scales and an unusual
logarithmic dependence of $\chi'$ on frequency $f$.  Crossing back
into the quantum paramagnet (D) restores independence to the annealing
history.}
  \label{QC3}
\end{figure}

Spectroscopy provides insight into the distribution and magnitude of
relaxation times for spin reorientation.  Fig.~\ref{QC3} shows
$\chi'(f)$ over four decades of frequency at various points A to D in
the phase diagram of Fig.~\ref{QC1}.  Spin relaxation proceeds
conventionally, independent of annealing protocol, in the classical
ferromagnet (Fig.~\ref{QC3}A).  As temperature is reduced to $T=0.4$~K
(Fig.~\ref{QC3}B), differences start to emerge.  The
classically-cooled spectrum is approximately half a decade to the left
of the quantum-cooled data; the quantum protocol has yielded a state
with relaxation times a factor of three smaller than its classical
counterpart.  This effect increases dramatically on cooling deep into
the ordered phase (Fig.~\ref{QC3}C).  As $f$ decreases toward 1~Hz,
$\chi'(f)$ grows logarithmically with roughly similar slopes for both
annealing histories.  The displacement between the two lines is now
one and one-half decades; quantum cooling has produced a state for
which the relaxation times are a factor of 30 faster than produced by
classical cooling.  Finally, keeping $T$ fixed at 0.030~K and
increasing $H_t$ back into the paramagnet (Fig.~\ref{QC3}D) narrows
the spectroscopic response and restores the system's insensitivity to
annealing method.

\begin{figure}[p!]
  \centering \epsfig{file=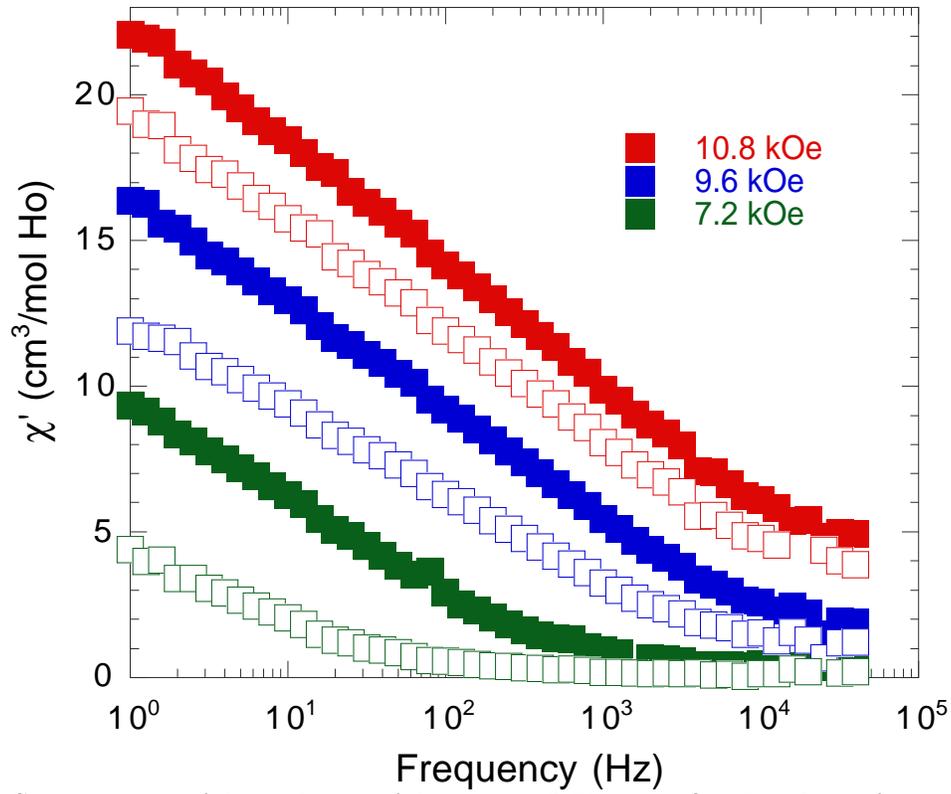, width=5in}
  \caption
[Spectroscopy of $\chi'(f)$ over five decades, demonstrating faster time
scales for quantum annealing.]
{Spectroscopy of the real part of the susceptibility over five decades
in frequency, demonstrating a faster time scale for quantum annealing
over its classical counterpart, as well as a well-defined logarithmic
divergence at low frequencies.}
  \label{QC4}
\end{figure}

The logarithmic divergence of $\chi'(f)$ at $T=0.030$~K is
shown (Fig.~\ref{QC4}) over nearly five decades in frequency.  As
$H_t$ grows, the spectra move to the right; there is a
spectacular acceleration of the relaxation.  At the same time, the
shift between the spectra obtained by quantum and classical cooling
algorithms shrinks so that at 10.8~kOe quantum cooling renormalizes
the relaxation times by only a factor of two.

\begin{figure}[p!]
  \centering \epsfig{file=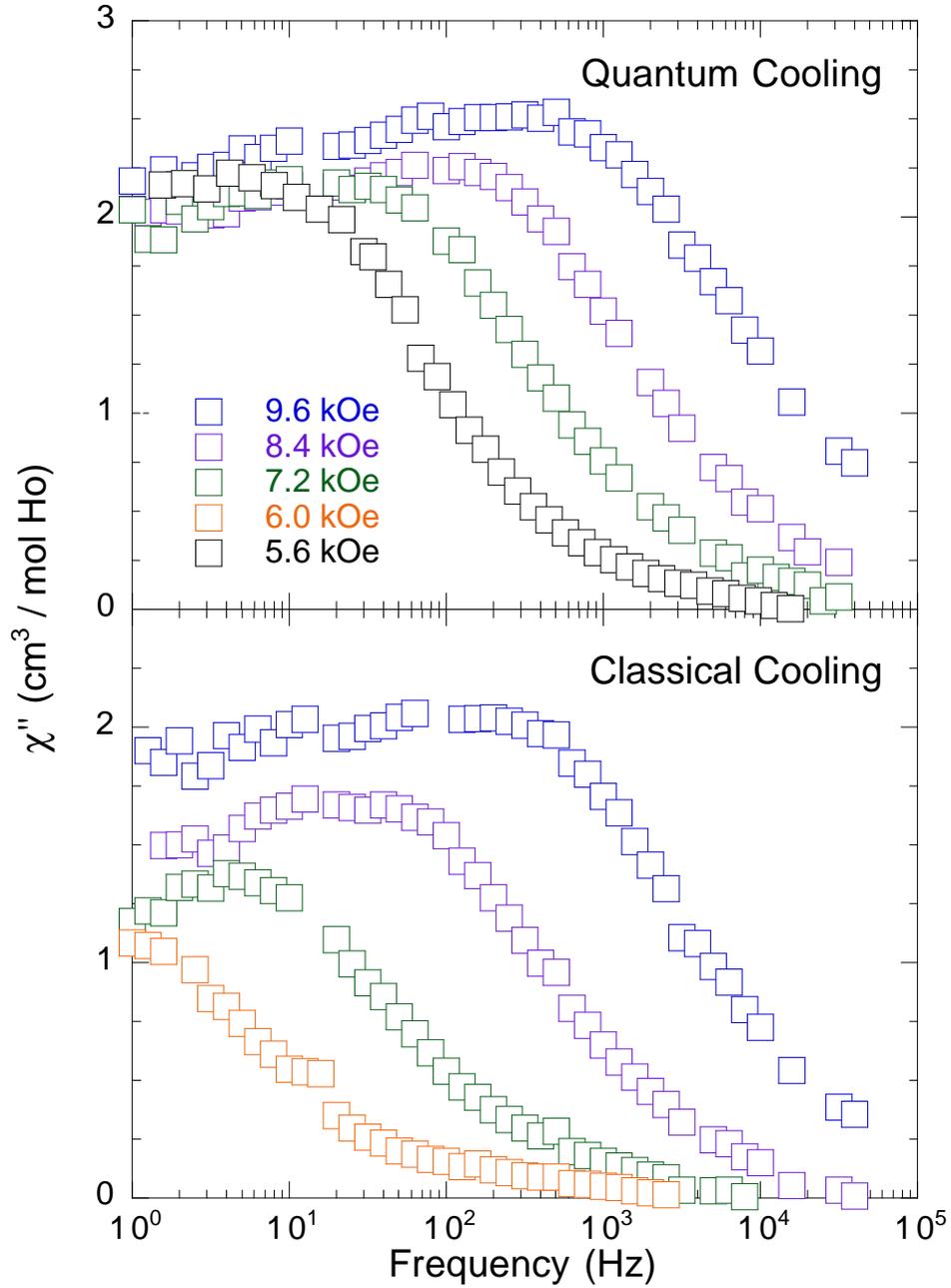, width=5in}
  \caption
[Imaginary susceptibility, $\chi''(f)$, highlighting differences between
quantum and classical protocols.]
{The imaginary component of the susceptibility, $\chi''(f)$, not only 
serves to emphasize the faster typical times when quantum tunneling is
featured, but reveals an apparent settling into the same state at low
frequency independent of transverse field endpoint.  The classical
computation does not appear to possess such simple convergent
properties.}
  \label{QC5}
\end{figure}

The different time scales for classical and quantum annealing are also
clearly evident in the imaginary part of the susceptibility,
$\chi''(f)$ (Fig.~\ref{QC5}), where the position of the peak response
is the inverse of a typical relaxation time.  Moreover, the low
frequency dissipative response provides an additional basis for
comparison.  The quantum annealed data all converge and flatten at low
frequencies, independent of transverse field destination at low $T$,
while the classical curves appear to head towards separate low
frequency, long time values.  It is natural to associate the
convergent state in the quantum computation with the ground state of
the system.  If the $\chi''(f,H_t)$ quantum response remains
coincident and frequency independent as $f \rightarrow 0$, then by the
Kramers-Kronig relation the real components, $\chi'(f,H_t)$, should be
logarithmic in $f$, in agreement with what was measured directly
(filled squares, Fig.~\ref{QC4}).  By the same token, the more diverse
low frequency behavior of $\chi''(f)$ after classical cooling is
reflected in the generally greater bowing and reduced parallelism
between the classical $\chi'(f)$ curves of Fig.~\ref{QC4}.

The logarithmically divergent $\chi'(f) \propto \ln{(f/f_\circ)}$,
which characterizes the state G in the phase diagram of Fig.~\ref{QC1}
implies that G is critical or marginally stable~\cite{BTW}.  It is a
nearly ferromagnetic state with equally probable fluctuations out of
that state on all (long) time scales.  Quantum and classical annealing
protocols yield states which differ primarily in the characteristic
frequency $f_\circ$.  The dramatically enhanced $f_\circ$ found for
quantum cooling shows that the addition of a tunneling term to the
Hamiltonian (Eq.~\ref{eq:Hamiltonian}) yields a state with more rapid
fluctuations.  Most importantly, quantum cooling has allowed us to see
more clearly and quickly that the ground state is likely a critical
state: at $T = 0. 030$~K and $H_t$ = 7.2~kOe, there is only one decade
of logarithmic behavior for classical cooling whereas there are
already two and a half decades for quantum cooling.

Our experiments on Li\sub{Ho}{\footnotesize 0.44}\sub{Y}{\footnotesize
0.56}\sub{F}{\footnotesize4} directly demonstrate the power of a
quantum mechanical term in the Hamiltonian for reaching a convergent
solution, with obvious implications for designing simulated annealing
computer algorithms~\cite{Kadowaki}.  They also raise more fundamental
issues about strategies for the design of actual quantum computers.
To date, the favored route has involved building a quantum computer
qubit by qubit, taking advantage of the extraordinary sensitivity and
isolation possible in modern nuclear magnetic resonance
techniques~\cite{Monroe,Gersh}.  Elaborate protocols for addressing
and manipulating individual qubits are then required to perform
computations.  This work allows speculation about a less refined
approach; cast the computation as a classical spin problem, which
is then solved by a combination of thermal cooling and the blanket
application of a transverse field.  The transverse field, which
controls the tunneling probability, is eventually driven to a low or
zero value, and the solution can be read out in the form of the
resulting and --- at that point --- frozen spin configuration

We have benefited greatly from discussions with P. Chandra, S. N. 
Coppersmith, and A. Ramirez.  The work at the University of Chicago 
was supported primarily by the Materials Research Science and Engineering 
Center (MRSEC) Program of the NSF under award number DMR-9808595.

\bibliography{lihof}
\bibliographystyle{prsty}

%
\end{document}